\documentclass[12pt]{article}

\begin{document}

\title{A conformally flat realistic anisotropic model for a compact star}
\author{B. V. Ivanov \\
Institute for Nuclear Research and Nuclear Energy, \\
Bulgarian Academy of Science, \\
Tzarigradsko Shausse 72, Sofia 1784, Bulgaria}
\maketitle

\begin{abstract}
A physically realistic stellar model with a simple expression for the energy
density and conformally flat interior is found. The relations between the
different conditions are used without graphic proofs. It may represnet a
real pulsar.
\end{abstract}

\section{Introduction}

The study of relativistic stellar structure is now more than 100 years old.
For a long time the star interior was considered to be made of perfect
fluid, which has equal radial $p_r$ and tangential $p_t$ pressures. This
leads to the isotropic condition $p_r=p_t$, imposed on the Einstein
equations. However, spherical symmetry demands only the equality of the two
tangential pressures.

In 1972 Ruderman \cite{one} argued for the first time that nuclear matter at
very high densities $\rho $ of the order of $10^{15}g/cm^3$ may have
anisotropic features and its interactions are relativistic. The pioneering
work of Bowers and Liang \cite{two} on building anisotropic models in 1974
gave start to a number of such solutions. Anisotropy may have a lot of
sources \cite{three}: a mixture of fluids of different types, presence of a
superfluid, existence of a solid core, phase transitions, presence of
magnetic field, viscosity, etc. Such models describe compact stellar objects
like neutron stars, strange stars, quark stars, boson stars, gravastars,
dark stars and others.

The Einstein equations describe the effect of matter upon the metric of
spacetime. For static, spherically symmetric fluid solutions the metric may
be written in comoving canonical coordinates and has two components $\nu $
and $\lambda $. The energy-momentum tensor is represented by its diagonal
components, mentioned above: $\rho $, $p_r$ and $p_t$. There are only three
equations for these five characteristics, so that two of them may be chosen
freely. They should satisfy, however, a lot of regularity, stability and
energy conditions for a realistic model. The situation with this
undetermined system of differential equations is analogous to the one for
charged isotropic star models \cite{four}. This is not surprising since
charge can be looked upon as an effective anisotropy of the model \cite{five}%
. Different choices of the two given functions have been made.

The simplest one is to propose ansatze for the two metric functions. One of
the first was given in \cite{six}, where some of the Tolman isotropic
solutions \cite{seven} were modified to become anisotropic. Other followed
recently \cite{eight}, \cite{nine}, \cite{ten}, \cite{eleven}, \cite{twelve}%
, \cite{thirteen}, \cite{fourteen}.

String theory has inspired embedding of branes like in the Randall - Sandrum
model \cite{fifteen}. This rekindled the interest in stellar models embedded
in five-dimensional flat spacetime (embedding class one). They must satisfy
the Karmarkar condition \cite{sixteen}. It can be written as a relation
between the metric functions and one of them can generate the whole
solution. It is interesting that the isotropic condition, can be translated
into a similar relation, giving different generating functions \cite
{seventeen}, \cite{eighteen}, \cite{nineteen}, \cite{twenty}, \cite{twone}.

There are just two perfect fluid solutions of the Karmarkar condition - the
interior Schwarzschild one and a cosmological one. When the fluid is
anisotropic, a number of realistic solutions has been found in the last two
years \cite{twtwo}, \cite{twthree}, \cite{twfour}, \cite{twfive}, \cite
{twsix}, \cite{twseven}, \cite{tweight}, \cite{twnine}, \cite{thirty}, \cite
{thone}, \cite{thtwo}, \cite{ththree}, \cite{thfour}, \cite{thfive}, \cite
{thsix}, \cite{thseven}, \cite{theight}, \cite{thnine}.

Conformally flat anisotropic spheres have vanishing Weyl tensor. This leads
to a differential equation for $\lambda $ and $\nu $, similar to the
Karmarkar one or the isotropic condition. Early solutions were found in \cite
{forty} where different ansatze for the mass function $m$ were proposed.
There is a simple relation between $m$ and $\lambda $ and from the condition
for conformal flatness $\nu $ may be found. Then expressions for all other
characteristics of the model are obtained. The first who integrated the
vanishing Weyl condition seems to be Ponce de Leon in 1987 \cite{foone}, but
no details were given. A relation between $\nu $ and $\lambda $ is the
outcome. Details were supplied in 2001 \cite{fotwo} and some models were
discussed with $p_r=0$ or prescribed $\lambda $. The authors worked in
non-comoving coordinates and gave the general solution in another, but
equivalent form, used later in \cite{fothree}. Conformally flat spherically
symmetric spacetimes were studied in different coordinate systems in \cite
{fofour}. A conformally flat model with polytropic equation of state was
discussed in \cite{fofive}. Other solutions were given too \cite{fosix}, 
\cite{foseven}.

Closely related are solutions which admit conformal motion. Some of them
depend on the conformal factor and a matter component, which can be $\rho $ 
\cite{foeight}, \cite{fonine}, or the mass $m$ \cite{fifty}. One can add
here models with given $\lambda $ \cite{fione}, \cite{fitwo}, since the
expressions for $\rho $, $m$ and $\lambda $ are simply related. There is a
model with a linear equation of state (LEOS) between $p_r$ and $\rho $ \cite
{fithree}, with two fluids \cite{fifour} and another one with LEOS between
the pressures \cite{fifive}. The work \cite{fotwo} has been generalized to
non-static solutions \cite{fosix}, \cite{foseven} and many new solutions
were obtained.

The main shortcoming of the existing model building is that the conditions
for a realistic model are checked after the ansatze for the two free
functions are made. The expressions for the different characteristics become
very involved even for polynomial seeding functions and one has to turn to
graphic proofs. Solutions usually have lots of constants in order to satisfy
the set C1-C10, introduced in Sect. 3. One constant turns a 2-dimensional
plot into a 3-dimensional one. With two and more constants only partial
plots are possible.

Recently, \cite{fisix} we have argued that the combination of free functions 
$\rho $, $p_r$ is the right choice to reduce the number of graphic proofs.
Another important fact is that the conditions C1-C10 are not independent.
There are many relations between them and we have reduced the set to a
couple of inequalities. Only they need in general a graphic proof in the
concrete examples. To illustrate this formalism we have given a solution
with simple energy density and linear EOS with bag constant.

In the present paper we apply the approach of \cite{fisix} to conformally
flat solutions with simple metric function $\lambda $. We make a full
analytic physical analysis of the solution and show that no graphic proofs
are necessary. It implies that a certain constant of the model should fall
in a particular range. A real pulsar is shown to satisfy this constraint.

In Sect. 2 the Einstein field equations are given, as well as the
definitions of the main characteristics of a static anisotropic star. The
Weyl condition and its solution are also introduced. In Sect. 3 we summarize
the conditions for a physically realistic model. In Sect. 4 we present the
model, which depends on three constants. In Sect. 5 we perform a full
physical analysis and find the range of the constants where C1-C10 are
fulfilled. In Sect. 6 a real star is shown to satisfy these constraints and
therefore is a candidate for a neutron star with conformally flat interior.
Sect. 7 contains a discussion.

\section{Field equations and definitions}

The interior of static spherically symmetric stars is described by the
canonical line element 
\begin{equation}
ds^2=e^\nu c^2dt^2-e^\lambda dr^2-r^2\left( d\theta ^2+\sin ^2\theta
d\varphi ^2\right) ,  \label{one}
\end{equation}
where $\lambda $ and $\nu $ are dimensionless and depend only on the radial
coordinate $r$. The Einstein equations read 
\begin{equation}
kc^2\rho =\frac 1{r^2}\left[ r\left( 1-e^{-\lambda }\right) \right] ^{\prime
},  \label{two}
\end{equation}

\begin{equation}
kp_r=-\frac 1{r^2}\left( 1-e^{-\lambda }\right) +\frac{\nu ^{\prime }}%
re^{-\lambda },  \label{three}
\end{equation}
\begin{equation}
kp_t=\frac{e^{-\lambda }}4\left( 2\nu ^{\prime \prime }+\nu ^{\prime 2}+%
\frac{2\nu ^{\prime }}r-\nu ^{\prime }\lambda ^{\prime }-\frac{2\lambda
^{\prime }}r\right) ,  \label{four}
\end{equation}
where $\rho $ is the matter density, $p_r$ is the radial pressure, $p_t$ is
the tangential one, $^{\prime }$ means a radial derivative and 
\begin{equation}
k=\frac{8\pi G}{c^4}.  \label{five}
\end{equation}
Here $G$ is the gravitational constant and $c$ is the speed of light.

The gravitational mass in a sphere of radius $r$ is given by 
\begin{equation}
m=\frac{kc^2}2\int_0^r\rho \left( \omega \right) \omega ^2d\omega .
\label{six}
\end{equation}
Due to $kc^2$, its dimension is length. Then Eq. (2) gives 
\begin{equation}
e^{-\lambda }=1-\frac{2m}r.  \label{seven}
\end{equation}
The compactness of the star $u$ is defined by 
\begin{equation}
u=\frac{2m}r  \label{eight}
\end{equation}
and is dimensionless.

On the other side, the redshift $Z$ depends on $\nu $: 
\begin{equation}
Z=e^{-\nu /2}-1.  \label{nine}
\end{equation}
The field equations do not contain $\nu $, but its first and second
derivative. One can express $\nu ^{\prime }$ from Eqs. (2), (3), and (7) as 
\begin{equation}
\nu ^{\prime }=\frac{krp_r+2m/r^2}{1-2m/r}.  \label{ten}
\end{equation}
The second derivative $\nu ^{\prime \prime }$ may be excluded by
differentiation of Eq. (3) and combination with the other field equations.
The result is 
\begin{equation}
p_r^{^{\prime }}=-\frac 12\left( \rho c^2+p_r\right) \nu ^{\prime }+\frac{%
2\Delta }r,  \label{eleven}
\end{equation}
where $\Delta =p_t-p_r$ is the anisotropic factor. Combining (10) and (11)
one gets the well-known TOV (Tolman, Oppenheimer, Volkoff) equation \cite
{seven}, \cite{fiseven} of hydrostatic equilibrium in a relativistic star,
found initially for isotropic solutions. Its anisotropic version was given
by Bowers and Liang \cite{two}: 
\begin{equation}
p_r^{^{\prime }}=-\left( \rho c^2+p_r\right) \frac{krp_r+2m/r^2}{2\left(
1-2m/r\right) }+\frac{2\left( p_t-p_r\right) }r.  \label{twelve}
\end{equation}
The hydrostatic force on the left $F_h$ is balanced by the gravitational $%
F_g $ and the anisotropic forces $F_a$ on the right. This equation is not
independent from the field equations, but is their consequence. It can
replace one of them. It is also equivalent to the Bianchi identities $T_{\nu
;\mu }^\mu =0$, which in the static spherically symmetric case have only one
non-trivial component \cite{two}, \cite{fieight}, \cite{finine}, \cite{sixty}%
. In CGS units $G=6.674\times 10^{-8}$ $cm^3/g.s^2$, $c=3\times 10^{10}$ $%
cm/s$, $k=2.071\times 10^{-48}$ $s^2/g.cm$, $kc^2=1.864\times 10^{-27}$ $%
cm/g $. The mass in grams $M$ is related to $m$ by 
\begin{equation}
m=\frac{GM}{c^2}.  \label{thirteen}
\end{equation}
From now on we set $G=c=1$, passing to usual relativistic units. Then $%
k=8\pi $. As a whole, we have three field equations for five unknown
functions: $\lambda ,\nu ,\rho ,p_r$ and $p_t$.

The space-time is conformally flat when its Weyl tensor vanishes. In our
case this gives a relation between the two metric coefficients \cite{foone} 
\begin{equation}
\frac{1-e^\lambda }{r^2}-\frac{\nu ^{\prime }\lambda ^{\prime }}4-\frac{\nu
^{\prime }-\lambda ^{\prime }}{2r}+\frac{\nu ^{\prime \prime }}2+\frac{\nu
^{\prime 2}}4=0.  \label{fourteen}
\end{equation}
Similar relations arise in embeddings of class 1 \cite{twfive}, or in the
case of isotropic pressure \cite{twone}. Eq. (14) may be integrated. It
appears that for the first time this was done in \cite{foone}, but no
details of the integration method were given. These were described later in 
\cite{fotwo}, \cite{fofour}, \cite{foseven}. The result can be written as 
\cite{fotwo} 
\begin{equation}
e^{\nu /2}=C_1r\cosh \left( \int \frac{e^{\lambda /2}}rdr+C\right) ,
\label{fifteen}
\end{equation}
where $C$ and $C_1$ are integration constants. This equation should be added
to the three field equations, so one may choose freely one generating
function to obtain solutions. The model will be physically realistic if a
number of regularity, matching and stability conditions are satisfied too.

\section{Conditions for a physically realistic model}

A comparatively reasonable set of conditions includes

C1. The metric potentials are positive and should be finite and free from
singularities in the star's interior and at the centre should satisfy $%
e^{\lambda \left( 0\right) }=1$ and $e^{\nu \left( 0\right) }=const$.

C2. Matching conditions. At the surface of the star $r=r_s$ the interior
solution should match continuously to the exterior Schwarzschild solution, 
\begin{equation}
ds^2=\left( 1-\frac{2m_s}{r_s}\right) dt^2-\left( 1-\frac{2m_s}{r_s}\right)
^{-1}dr^2-r^2\left( d\theta ^2+\sin ^2\theta d\varphi ^2\right) ,
\label{sixteen}
\end{equation}
where $m_s=m\left( r_s\right) $. This determines the metric at the surface 
\begin{equation}
e^{\nu \left( r_s\right) }=e^{-\lambda \left( r_s\right) }=1-\frac{2m_s}{r_s}%
.  \label{seventeen}
\end{equation}
In addition, the radial pressure there vanishes, $p_{rs}=0$. Neither the
energy density nor the tangential pressure are obliged to do so.

C3. The interior redshift $Z$, given by Eq. (9), should decrease with the
increase of $r$. The surface redshift and compactness are related, due to
Eq. (17): 
\begin{equation}
Z_s=\left( 1-u_s\right) ^{-1/2}-1.  \label{eighteen}
\end{equation}
They should be less than the universal bounds, found when different energy
conditions hold (see C6). In the isotropic case they are $2$ and $8/9$
correspondingly \cite{sione}. In the anisotropic case, when DEC holds, they
are $5.211$ and $0.974$. When TEC holds, one has the bounds $3.842$ and $%
0.957$ \cite{sitwo}. They are greater than those in the isotropic case, but
not arbitrary as asserted in \cite{two}.

C4. The density and the pressures should be non-negative inside the star. At
the centre they should be finite $\rho \left( 0\right) =\rho _0$, $p_r\left(
0\right) =p_{r0}$, $p_t\left( 0\right) =p_{t0}$. Moreover, $p_{r0}=p_{t0}$ 
\cite{sitwo}

C5. They should reach a maximum at the centre, so $\rho ^{\prime }\left(
0\right) =p_r^{\prime }\left( 0\right) =p_t^{\prime }\left( 0\right) =0$ and
should decrease monotonically outwards, $\rho ^{\prime }\leq 0$, $%
p_r^{\prime }\leq 0$, $p_t^{\prime }\leq 0$. The tangential pressure should
remain bigger than the radial one, except at the centre, $p_t\geq p_r$.

C6. Energy conditions. The solution should satisfy the dominant energy
condition (DEC) $\rho \geq p_r$, and $\rho \geq p_t$. The strong energy
condition (SEC) \cite{sithree} should be satisfied too, $\rho +p_r+2p_t\geq
0.$ Because of C4 it is trivial. It is desirable that even the trace energy
condition (TEC) $\rho \geq p_r+2p_t$ should be satisfied. Obviously, the
latter is stronger than DEC.

C7. Causality condition. It says that the radial and tangential speeds of
sound should not surpass the speed of light. The speeds of sound are defined
as $v_r^2=dp_r/d\rho $ and $v_t^2=dp_t/d\rho $. Therefore this condition
reads 
\begin{equation}
0<\frac{dp_r}{d\rho }\leq 1,\quad 0<\frac{dp_t}{d\rho }\leq 1.
\label{nineteen}
\end{equation}

C8. The adiabatic index $\Gamma $ as a criterion of stability. This index is
the ratio of two specific heats and should be bigger than $4/3$ for
stability \cite{three}, \cite{sifour}, \cite{sifive}, 
\begin{equation}
\Gamma =\frac{\rho +p_r}{p_r}\frac{dp_r}{d\rho }\geq \frac 43.
\label{twenty}
\end{equation}

C9. Stability against cracking. Cracking was introduced by Herrera \cite
{sisix} as a possibility of breaking of perturbed self-gravitating spheres.
Abreu et al \cite{siseven} found a simple requirement for avoiding this to
happen, namely the region of stability is 
\begin{equation}
-1\leq \frac{dp_t}{d\rho }-\frac{dp_r}{d\rho }\leq 0.  \label{twone}
\end{equation}

C10. The Harrison-Zeldovich-Novikov stability condition \cite{sieight}, \cite
{sinine}. It implies that $dM\left( \rho _0\right) /d\rho _0>0$.

\section{The model}

We shall choose a simple ansatz for $e^\lambda $ as a generating function,
namely 
\begin{equation}
e^{-\lambda }=\left( 1-x\right) ^2,\quad x=\frac{r^2}{b^2},  \label{twtwo}
\end{equation}
where $b$ is some constant of dimension length, so that $x$ is
dimensionless, as is the metric coefficient. Its range is from $0$ to $x_s<1$%
.Then Eqs. (7) and (8) give 
\begin{equation}
m=\frac 12r\left( 2x-x^2\right) ,\quad u=2x-x^2.  \label{twthree}
\end{equation}
The derivative of Eq. (6) yields 
\begin{equation}
k\rho =\frac{2m^{\prime }}{r^2}  \label{twfour}
\end{equation}
and using this formula or Eq. (2) we obtain for the energy density 
\begin{equation}
kb^2\rho =6-5x,  \label{twfive}
\end{equation}
which is very simple. In more general form it was used in the past, \cite
{fisix}, \cite{seventy}, \cite{seone}, \cite{setwo}, \cite{sethree}, \cite
{sefour}, \cite{sefive}, \cite{sesix}, \cite{seseven}, \cite{seeight}, \cite
{senine}, \cite{eighty}, \cite{eione}. In the context of conformal flatness
it was used in \cite{forty}, Example 4, and \cite{fotwo}, model III but only
a partial physical analysis has been done. Eq. (25) clarifies the meaning of 
$b$: 
\begin{equation}
b^2=\frac 6{k\rho _0}.  \label{twsix}
\end{equation}
The zero index will be used for variables at the centre of the star. Thus $b$
is related to the central density $\rho _0$, whose value in CGS units is
about $10^{15}g$ for compact neutron stars.

Let us introduce now the constants $B$ and $\alpha $ instead of $C$%
\begin{equation}
C=\frac 12\ln B^2,\quad \alpha =B^2-1.  \label{twseven}
\end{equation}
Then Eq. (15) gives an expression for the other metric coefficient 
\begin{equation}
e^\nu =\frac{C_1^2b^2}{4\left( 1+\alpha \right) }\frac{\left( 1+\alpha
x\right) ^2}{1-x},  \label{tweight}
\end{equation}
which is obviously positive. The redshift $Z$ throughout the star is
obtained then from Eq. (9). The derivative of $\nu $ with respect to $x$,
which enters the field equations, is 
\begin{equation}
\nu _x=\frac{2\alpha }{1+\alpha x}+\frac 1{1-x}  \label{twnine}
\end{equation}
and is positive too. Eq. (3), which is an expression for the radial
pressure, yields the formula 
\begin{equation}
kb^2p_r=-x+\frac{4\alpha \left( 1-x\right) ^2}{1+\alpha x}.  \label{thirty}
\end{equation}

There is a general expression for the anisotropy factor $\Delta $ in
conformally flat models, which follows from the field equations (3) and (4)
and the requirement (14) \cite{fothree}, \cite{sitwo} 
\begin{equation}
\frac{k\Delta }r=-2\left( \frac m{r^3}\right) ^{\prime }.  \label{thone}
\end{equation}
In the case of the simple ansatz (22) it becomes 
\begin{equation}
kb^2\Delta =2x,\quad kb^2\Delta _x=2.  \label{thtwo}
\end{equation}
It makes the expression for the tangential pressure very similar to the one
for the radial pressure 
\begin{equation}
kb^2p_t=x+\frac{4\alpha \left( 1-x\right) ^2}{1+\alpha x}.  \label{ththree}
\end{equation}
Thus, the characteristics of the model are given by simple elementary
functions. They depend on three constants $b$ (or $\rho _0$), $\alpha $ and $%
C_1$. They should be related to the mass $m_s$ and the radius $r_s$ of the
star.

\section{Physical analysis}

\quad Now we have to choose the free parameters of the model in such a way
that the conditions C1-C10 are satisfied.

C1. Eq. (22) shows that $e^\lambda $ is finite and positive and increases
monotonically with $r$ from $1$ to $\left( 1-x_s\right) ^{-2}$. Eqs. (28)
and (29) show that $e^\nu $ is also finite and positive and increases
monotonically. Eq. (17) shows that $e^{\nu \left( r_s\right) }$ is less than
1, hence $e^{\nu \left( 0\right) }$ is also less than one.

C2. The matching condition for $\lambda $ is fulfilled when $m_s=m\left(
r_s\right) $. Thus Eq. (23) gives 
\begin{equation}
m_s=\frac b2x_s^{3/2}\left( 2-x_s\right) ,\quad u_s=2x_s-x_s^2.
\label{thfour}
\end{equation}
Eqs. (17), (22) and (28) fix $C_1$%
\begin{equation}
C_1^2=\frac{4\left( 1+\alpha \right) \left( 1-x_s\right) ^3}{b^2\left(
1+\alpha x_s\right) ^2}  \label{thfive}
\end{equation}
in terms of $\alpha $, $b$ and $r_s$. The boundary condition $p_{rs}=0$,
combined with Eq. (30) expresses $\alpha $ as a function of $x_s$%
\begin{equation}
\alpha =\frac{x_s}{\left( 2-x_s\right) \left( 2-3x_s\right) }.  \label{thsix}
\end{equation}
Thus $\alpha $ is positive, increases monotonically with $x_s$ and is finite
as long as $x_s<2/3$. Then $B^2$ is positive as it should be.

C3. Eqs. (9) and (28) show that $Z$ decreases monotonically throughout the
star and at the surface 
\begin{equation}
Z_s=\left( 1-x_s\right) ^{-1}-1,  \label{thseven}
\end{equation}
due to Eqs. (18) and (34).

C4. Because of Eq. (25) $\rho $ will be positive as long as $x<6/5.$ This
inequality is true because $x_s<2/3$. The energy-density is finite at the
centre and taken to be about $10^{15}g/cm^3$. This defines $b$ according to
Eq. (26). Both terms in the expression for $p_r$ in Eq. (30) decrease
monotonically when $r$ increases. We have arranged that $p_{rs}=0$ (Eq. 36).
Hence, in the interior $p_r$ decreases monotonically to zero and is
positive. This is confirmed by its derivative 
\begin{equation}
kb^2p_{rx}=-1-\frac{4\alpha \left( 1-x\right) \left( 2+\alpha +\alpha
x\right) }{\left( 1+\alpha x\right) ^2}.  \label{theight}
\end{equation}
It is obviously negative and since $x^{\prime }=2r/b^2$, $p_r^{\prime }$ is
also negative. Finally, due to Eq. (32), $\Delta =p_t-p_r\geq 0$ and $%
p_t\geq p_r$. Therefore $p_t$ is also positive and at $r=0$ coincides with $%
p_r$. Their value at the centre is given by Eq. (30) 
\begin{equation}
p_{t0}=p_{r0}=\frac{4\alpha }{kb^2}.  \label{thnine}
\end{equation}

C5. Eq. (25) gives 
\begin{equation}
kb^2\rho _x=-5.  \label{forty}
\end{equation}
Eqs. (32), (38) and (40) combine to deliver $\rho ^{\prime }\left( 0\right)
=p_r^{\prime }\left( 0\right) =p_t^{\prime }\left( 0\right) =0$ and the
monotonic decrease of $\rho $ and $p_r$. It remains to prove that $%
p_t^{\prime }\leq 0$. Before doing that let us turn to C7.

C7a. Causality condition for $dp_r/d\rho $. This ratio can be written as $%
p_{rx}/\rho _x$. We have just proved that the numerator and the denominator
are negative, so their ratio is positive. Hence, the left inequality of the
first part of Eq. (19) is true. The right inequality demands 
\begin{equation}
-kb^2p_{rx}\leq 5,  \label{foone}
\end{equation}
because of Eq. (40). Let us go now to C9.

C9. The anti-cracking condition may be written as 
\begin{equation}
-1+\frac{dp_r}{d\rho }\leq \frac{p_t^{\prime }}{\rho ^{\prime }}\leq \frac{%
p_r^{\prime }}{\rho ^{\prime }}.  \label{fotwo}
\end{equation}
We suppose that inequality (41) holds. Then the left hand side of Eq. (42)
is negative. Let us multiply Eq. (42) by $\rho ^{\prime }$, which was shown
to be negative. We have 
\begin{equation}
\rho ^{\prime }\left( -1+\frac{dp_r}{d\rho }\right) \geq p_t^{^{\prime
}}\geq p_r^{^{\prime }}.  \label{fothree}
\end{equation}
Now the left hand side is positive. Combining this chain of inequalities
with the inequality $p_t^{\prime }\leq 0$, that we have to prove, we obtain 
\begin{equation}
0\geq p_t^{^{\prime }}\geq p_r^{^{\prime }}.  \label{fofour}
\end{equation}
Then we shall finish the proof of C5 and C9. These are the same
inequalities, derived in \cite{fisix}, Eq. (26). In addition, since Eq. (44)
may be written as 
\begin{equation}
0\leq \frac{dp_t}{d\rho }\leq \frac{dp_r}{d\rho }\leq 1,  \label{fofive}
\end{equation}
we also prove C7b, the causality condition for $dp_t/d\rho $. Thus, C5 about 
$p_t^{\prime }$, C7 and C9 are reduced to Eqs. (41) and (44). Moving to $x$%
-derivatives and subtracting $p_{rx}$ from Eq. (44) we obtain 
\begin{equation}
-p_{rx}\geq \Delta _x\geq 0.  \label{fosix}
\end{equation}
Utilizing Eq. (32), we transform the above two inequalities into one: 
\begin{equation}
-kb^2p_{rx}\geq 2.  \label{foseven}
\end{equation}
Finally, Eqs. (41) and (47) yield 
\begin{equation}
2\leq -kb^2p_{rx}\leq 5.  \label{foeight}
\end{equation}

To solve these two inequalities we use Eq. (38). Then the left inequality
becomes 
\begin{equation}
5\alpha ^2x^2+10\alpha x+1-4\alpha \left( 2+\alpha \right) \leq 0,
\label{fonine}
\end{equation}
while the right inequality transforms into 
\begin{equation}
2\alpha ^2x^2+4\alpha x+1-\alpha \left( 2+\alpha \right) \geq 0.
\label{fifty}
\end{equation}
The terms containing $x$ in Eq. (50) increase with $x$, hence, it is enough
to prove it for $x=0$. Then it becomes an inequality for $\alpha $%
\begin{equation}
\alpha ^2+2\alpha -1\leq 0.  \label{fione}
\end{equation}
The l.h.s. increases with $\alpha $ starting from $-1$, therefore $\alpha $
should be less or equal than the positive root of the corresponding equation 
\begin{equation}
\alpha \leq \alpha _1=-1+\sqrt{2}=0.414.  \label{fitwo}
\end{equation}
Eq. (36) is quadratic for $x_s$ with $\alpha $ as parameter. The root less
than one should be used to express $x_s$, namely 
\begin{equation}
x_{s1}=\frac{8\alpha +1-\sqrt{16\alpha ^2+16\alpha +1}}{6\alpha }.
\label{fithree}
\end{equation}
As we mentioned after Eq. (36), $x_s$ decreases with $\alpha $, hence $%
x_s\leq x_{s1}\left( \alpha _1\right) $ or 
\begin{equation}
x_s\leq \frac{8\sqrt{2}-7-\sqrt{33-16\sqrt{2}}}{6\left( \sqrt{2}-1\right) }%
=0.439.  \label{fifour}
\end{equation}

The terms containing $x$ in Eq. (49) increase with $x$, hence, it is enough
to prove it for $x_s$. Then, due to Eq. (36), it becomes a fourth degree
equation for $x_s$. Going back to Eq. (38), one can write Eq. (49) as 
\begin{equation}
\left( 1+\alpha x_s\right) ^2\leq 4\alpha \left( 1-x_s\right) \left(
2+\alpha +\alpha x_s\right) .  \label{fifive}
\end{equation}
Replacing $\alpha $ with its expression from Eq. (36) we get 
\begin{equation}
4\left( 1-x_s\right) ^4\leq x_s\left( 1-x_s\right) ^2\left( 8-7x_s\right) .
\label{fisix}
\end{equation}
The fourth degree inequality surprisingly becomes a quadratic one, when we
divide both sides by the common multiplier, and thus much easier to be
solved. We have 
\begin{equation}
11x_s^2-16x_s+4\leq 0.  \label{fiseven}
\end{equation}
The derivative of the l.h.s. is negative, so it decreases from 4 and becomes
negative at the point, where the inequality becomes an equality. We solve
this quadratic equation and take the root that is less than $1$. The
solution reads 
\begin{equation}
x_s\geq \frac{8-2\sqrt{5}}{11}=0.321.  \label{fieight}
\end{equation}
Combining Eqs. (54) and (58) we obtain the range of $x_s$%
\begin{equation}
0.321=\frac{8-2\sqrt{5}}{11}\leq x_s\leq \frac{8\sqrt{2}-7-\sqrt{33-16\sqrt{2%
}}}{6\left( \sqrt{2}-1\right) }=0.439.  \label{finine}
\end{equation}
In this range C5, C7 and C9 hold.

Let us discuss next the energy conditions C6. The left part of the proven
Eq. (19) may be written as 
\begin{equation}
\left( p_r-\rho \right) ^{\prime }\geq 0,  \label{sixty}
\end{equation}
since $\rho ^{\prime }\leq 0$. A definite integral of the l.h.s. is also
positive, 
\begin{equation}
\int_r^{r_s}\left( p_r-\rho \right) ^{\prime }dr=-p_r-\rho _s+\rho \geq 0,
\label{sione}
\end{equation}
which proves DEC for $p_r$, because $\rho _s\geq 0$.

The r.h.s. of Eq. (19) may be written as 
\begin{equation}
\left( p_t-\rho \right) ^{\prime }\geq 0  \label{sitwo}
\end{equation}
and the same integral of this inequality gives 
\begin{equation}
\rho -p_t\geq \rho _s-p_{ts}.  \label{sithree}
\end{equation}
Hence DEC for $p_t$ holds in the interior, if it holds at the surface of the
star. In \cite{fisix} a sufficient universal condition was given, $u_s\leq
0.8$. For our model Eqs. (34) and (59) give 
\begin{equation}
0.539\leq u_s\leq 0.685,  \label{sifour}
\end{equation}
so that the sufficient condition is satisfied in the whole range of $x_s$.
We can also use the expressions for $\rho $ (Eq. 25) and $p_t$ (Eq. (33)) to
find 
\begin{equation}
kb^2\left( \rho _s-p_{ts}\right) =2\left( 1-x_s\right) \left( 3-2\alpha
+5\alpha x_s\right) .  \label{sifive}
\end{equation}
Eqs. (36) and (59) show that $\alpha \in [0.184,0.414]$. In this interval $%
3-2\alpha $ is positive and consequently the r.h.s. of the above equation is
positive too. This proves that DEC holds for $p_t$ as well.

Let us prove, finally, that TEC is true. Eq. (61) gives $\rho \geq p_r+\rho
_s$. If 
\begin{equation}
\rho _s\geq 2p_{r0},  \label{sisix}
\end{equation}
TEC follows from the chain of inequalities 
\begin{equation}
\rho \geq p_r+2p_{r0}=p_r+2p_{t0}\geq p_r+2p_t.  \label{siseven}
\end{equation}
This chain is true due to Eq. (39) (the two pressures are equal at the
centre of the star) and Eq. (44) (the tangential pressure decreases towards
the stellar surface). Using Eqs. (25) and (30), Eq. (66) becomes 
\begin{equation}
6\geq 5x_s+8\alpha .  \label{sieight}
\end{equation}
As we know $\alpha $ increases with $x_s$ till $0.414$. Inserting this and
the maximum of $x_s=$ $0.439$ in the above inequality, yields $6\geq 5.567$,
which obviously is true. Hence, TEC holds for the whole range of $x_s$. Thus
C6 holds in its entirety.

Condition C8. In \cite{fisix} a sufficient condition was given for Eq. (20)
to hold, namely TEC and a lower limit for the radial speed of sound 
\begin{equation}
\frac{dp_r}{d\rho }\geq \frac 13.  \label{sinine}
\end{equation}
For our model TEC holds and 
\begin{equation}
\frac{dp_r}{d\rho }=\frac{kb^2p_{rx}}{kb^2\rho _x}=-\frac 15kb^2p_{rx},
\label{seventy}
\end{equation}
so that Eq. (69) becomes 
\begin{equation}
-kb^2p_{rx}\geq \frac 53.  \label{seone}
\end{equation}
This is true because of Eq. (48). Hence, the range of $x_s$, given by Eq.
(59) allows to prove that C1-C9 are true.

The final condition C10. Combining Eqs. (26) and (34) yields 
\begin{equation}
6m_s=k\rho _0r_s^3-\frac 1{12}k^2\rho _0^2r_s^5.  \label{setwo}
\end{equation}
The derivative of the total stellar mass with respect to the central
density, when the star radius is kept constant, reads 
\begin{equation}
\frac{dm_s\left( \rho _0\right) }{d\rho _0}=\frac 16kr_s^3\left( 1-x_s\right)
\label{sethree}
\end{equation}
and is obviously positive. Thus C10 is true and the whole set C1-C10 is true
as long as $x_s$ belongs to the range given by Eq. (59), $x_s\in
[0.321,0.439]$.

\section{Model of a real star}

The astronomers collect data about the radius $r_s$ and the mass $M$ of real
stars. Usually the ratio $\beta $ to the solar mass $M_{sol}$ is used 
\begin{equation}
\frac M{M_{sol}}=\frac{m_s}{m_{sol}}\equiv \beta .  \label{sefour}
\end{equation}
It is known that in relativistic units $m_{sol}=1.474$ km. Then we can find
the compactness from Eq. (8) 
\begin{equation}
u_s=2m_{sol}\frac \beta {r_s}=2.948\frac \beta {r_s},  \label{sefive}
\end{equation}
where $r_s$ is in km. The conformally flat solution is physically realistic
when $x_s\in [0.321,$ $0.439]$. Eqs. (7) and (22) or Eq. (23) give a
relation between $u_s$ and $x_s$%
\begin{equation}
x_s=1-\sqrt{1-u_s}.  \label{sesix}
\end{equation}
Eq. (64) may be used too, $u_s\in [0.539,$ $0.685]$. Then 
\begin{equation}
0.183\leq \frac \beta {r_s}\leq 0.232.  \label{seseven}
\end{equation}
Thus, it is very easy to find whether a real star may be described by our
model. Only the surface compactness of the star matters or equivalently, its
surface redshift. Eq. (37) gives the limits $Z_s\in [0.473,$ $0.782]$. These
are somewhat higher than the redshifts of many other models, discussed in
the literature.

An important characteristic is the central density $\rho _0$. We suppose
that it may be written as 
\begin{equation}
\rho _0=a\times 10^{15}\ g/cm^3,  \label{seeight}
\end{equation}
where $a$ is some constant close to 1. Eqs. (26) and (22) in CGS units give 
\begin{equation}
\rho _0=\frac{6x_s}{kc^2r_s^2}.  \label{senine}
\end{equation}
Thus $a$ is given by 
\begin{equation}
a=\frac{6\times 10^{-15}x_s}{kc^2r_s^2}=321.9\frac{x_s}{r_s^2},
\label{eighty}
\end{equation}
where we have used $kc^2=1.864\times 10^{-27}$ $cm/g$ and $r_s$ is given in
km.

Let us apply these formulas to some stars, described in a recent paper \cite
{eitwo}. There a model with given $\lambda $ and $\Delta $ was used. The
pulsar 4U1820-30 has $\beta =1.58$ and $r_s=9.1$ $km$. Then $\beta
/r_s=0.174 $ and is out of range. The pulsar Cen X-3 has $\beta =1.49$ and $%
r_s=9.178$ $km$. Again $\beta /r_s=0.162$ is too small. However, the pulsar
PSR J1614-2230 has $\beta =1.97$ and $r_s=9.69$ $km$ and $\beta /r_s=0.203$
which satisfies Eq. (77) and may be a candidate for a compact neutron star
with a conformally flat interior. We get from Eq. (75) $u_s=0.598$. Eq. (76)
yields $x_s=0.366$. Then we obtain from Eq. (80) $a=1.258$, which is
realistic. The surface redshift is comparatively high, $Z_s=0.577$ (see Eq.
(18) or (37)), but is less than the known limits \cite{sione}, \cite{sitwo}.
The other characteristics of this star may be found from the formulas in the
previous sections.

\section{Discussion}

We have followed in this paper the approach of \cite{fisix}. Although,
instead of $\rho $ and $p_r$, we chose an ansatz for $\lambda $ and the
condition of conformal flatness, we have been able to satisfy all physical
conditions without using graphic proofs. After all, C1-C10 involve
inequalities, which, in principle, may be proved by algebra and calculus.
Their use is limited to solving quadratic equations and integrating
derivatives. The relations between the different conditions, that we found
in the above reference allowed us to come to the same basic couple of
inequalities, Eq (44). It is interesting that the sufficient conditions for
the other realistic features of the model are contained in them and do not
restrict further the range of the main parameter $x_s$. A remarkable fact is
that only the compactness (or the surface redshift) of the star is necessary
to determine, whether it may have a conformally flat interior. The mass and
the radius of the star are enough to determine all of its characteristics.

In the perfect fluid case the Buchdahl bounds \cite{sione} on the
compactness and the redshift are $8/9=0.889$ and $2$ (see C3). They are
saturated by the Schwarzschild interior solution, which is an incompressible
sphere with constant density. It is unphysical, because the speed of sound
is infinite. Something more, the saturation occurs when the pressure is
infinite at the centre \cite{sitwo}. This model is the unique conformally
flat one for perfect (isotropic) fluids. This is one of the reasons to study
conformally flat solutions in the anisotropic case. It may provide an
explanation for the intermediate ranges of compactness and redshifts of
realistic anisotropic solutions.

In the literature, in many papers the real strong energy condition (SEC) is
used, which, however, is trivial. In many others, the trace energy condition
(TEC) is called SEC and made use of. It is really strong, because it
requires that the energy density (which in CGS units is multiplied by $c^2$)
should be bigger than the sum of the radial and the two equal tangential
pressures. Thus, it is even stronger than the dominant energy condition
(DEC). We have tried to clarify this misuse of notation.

Finally, the proofs of C1-C10 were considerably simplified by the simple
expression for the anisotropy factor $\Delta $. In the case of embeddings of
class one, the Karmarkar condition leads to a more sophisticated form for $%
\Delta $. Therefore, the present paper may be considered also as a
preparation to attack this case.

\end{document}